\documentclass[aps,prl,onecolumn,floatfix,tightenlines,showpacs,longbibliography,notitlepage]{revtex4-1}
\usepackage{amsfonts}
\usepackage{color}
\usepackage{amsmath}    
\usepackage{epsfig}
\usepackage{graphicx}
\usepackage{dcolumn}
\usepackage{bm}
\usepackage[breaklinks,colorlinks = true,linkcolor = blue,urlcolor=blue,citecolor=blue]{hyperref}
\usepackage{array}
\usepackage{url} 
\usepackage{lineno}
\usepackage{graphicx}
\usepackage{amsmath,amssymb}
\usepackage[nice]{nicefrac}

\newcommand{\ET}{E_{\mathrm{T}}}
\newcommand{\EP}{E_{\mathrm{P}}}
\newcommand{\pT}{p_{\mathrm{T}}}
\newcommand{\pP}{p_{\mathrm{P}}}
\begin{document}
\title{Lagrangian Irreversibility in Stratified Turbulence}
\author{Mrinal Jyoti Powdel}
 \email{mrinal.jyoti@icts.res.in}
\affiliation{International Centre for Theoretical Sciences, Tata Institute of Fundamental Research, Bengaluru 560089, India}
\author{Samriddhi Sankar Ray}%
 \email{samriddhisankarray@gmail.com}
\affiliation{International Centre for Theoretical Sciences, Tata Institute of Fundamental Research, Bengaluru 560089, India}
\begin{abstract}
Stable stratification reorganizes turbulence, suppressing vertical motion and
	altering the balance between vortical and strain-dominated flow
	structures. How these changes affect the temporal asymmetry of
	kinetic-energy exchange along individual fluid-particle trajectories
	remains incompletely understood. Here, we use direct numerical
	simulations of homogeneous Boussinesq turbulence to investigate
	Lagrangian power statistics across four classes of local
	velocity-gradient topology and three nonzero stratification strengths,
	with an unstratified reference case for comparison. The global power
	variance decreases with increasing stratification, while the
	standardized skewness becomes more negative, indicating enhanced
	negative asymmetry despite weaker overall fluctuations. Conditioning on
	local topology reveals distinct responses: axial-strain-dominated
	regions exhibit the most negative total-power skewness across all
	sampled stratification strengths, whereas biaxial-strain-dominated
	regions change from positive skewness in the unstratified reference
	case to negative skewness under stratification. Horizontal and vertical
	component-power statistics also respond differently, highlighting the
	directional dependence of the energy-exchange dynamics. An analysis of the dynamical contributions to power reveals that the pressure contribution, which is nearly symmetric in the unstratified case, becomes progressively more negatively skewed with an increase in stratification.
    Energy-conditioned pressure statistics further
	indicate changes in pressure-mediated energy redistribution as stratification increases. Finally, we find that the persistence-time distributions of the tracers exhibit exponential tails, with characteristic time scales that differ across topology classes and respond differently to stratification.
    These results show that increasing stratification is accompanied by three concurrent changes: an enhanced asymmetry of Lagrangian power, a modified redistribution of energy by pressure, and an altered temporal sampling of local flow structures by the tracers. Together, these co-occurring signatures characterize the response of the system to stratification, whereas the causal relations among them remain to be established.

\end{abstract}
\maketitle

\section{Introduction}

Stable density stratification is a defining feature of the ocean interior and
many atmospheric flows \cite{Davidson_2013, rileyannurev, staquetannurev,
caulfield_annurev, Caulfield_open_question}. By resisting vertical
displacement, buoyancy suppresses vertical motion and alters the balance
between horizontal and vertical dynamics. As a result, stratified turbulence
becomes anisotropic, with wave-like motions coexisting with vortical structures
and regions of strong deformation. These changes affect how turbulent energy is
distributed across scales and how, for example, momentum is transported
in geophysical flows \cite{Linden01011979, Riley1981, Fernando_1988,
Itsweire_Helland_1989, Ruddick1989597, Park_Whitehead_Gnanadeskian_1994,
Nicolleau_Vassilicos_2000, Chomaz_self_similarity, Dubrulle_transition,
Lindborg_Brethouwer_2008, Augier_Galtier_Billant_2012, Kimura_Herring_2012,
rorai_mininni_largescale, Herbert_Marino_Rosenberg_Pouquet_2016,
glazunov2019layered, Maffioli_sign, Labarre2024,
Varanasi2025,powdeluncertainty2026}.
Turbulence is commonly characterized through the statistical
properties of velocity fluctuations, providing an Eulerian description of the
flow at fixed locations. An equally important, complementary perspective is
the Lagrangian description, which follows individual fluid parcels as they
move through the flow. While the Eulerian approach characterizes the
statistical properties of the flow field, the Lagrangian approach tracks the
changes experienced by moving fluid parcels and is particularly relevant to
processes such as mixing and the transport of heat, salt, and other tracers.

Energy provides a natural link between the Eulerian and
Lagrangian perspectives. Turbulent flows maintained by forcing and
dissipation are paradigmatic nonequilibrium systems, in which energy is
transferred across scales and dissipated at small scales. Understanding
their dynamics therefore involves not only characterizing the distribution
of energy, but also examining how energy changes and is exchanged over
time. In this context, the statistical asymmetry of energy exchange along
individual fluid trajectories provides a way to investigate one aspect of
the temporal irreversibility of turbulent dynamics.

A Lagrangian description follows individual fluid parcels as they move
through a flow \cite{toschiannrev2009lagrangian}. Along a trajectory, the
instantaneous Lagrangian power $p=dE/dt$, where $E$ is the parcel's
kinetic energy, measures the rate of kinetic-energy change. Positive and
negative power correspond to energy gain and loss, respectively, so the
power distribution describes the statistical character of these exchanges
\cite{xuflightcrashPNAS2014, PumirredistributionPRX2014,
PumirPRL2016vortexstretching, Ray2018,picardo2020lagrangian}. In homogeneous and
isotropic, statistically stationary turbulent flows, this distribution is
negatively skewed: relatively gradual energy gains are followed by sharper
energy-loss events, a pattern referred to as ``flight--crash''
dynamics \cite{xuflightcrashPNAS2014}. The skewness
quantifies this asymmetry, with negative values indicating a more
pronounced tail toward negative power and thus a statistical imbalance
between energy-gain and energy-loss events. It therefore provides a
statistical diagnostic of the asymmetry of Lagrangian kinetic-energy
exchange.

This perspective is particularly relevant to stratified
turbulence, where buoyancy constrains vertical motion and reorganizes the
structures through which energy is transferred and redistributed. In the
ocean interior, these dynamics contribute to the mixing of heat, salt, and
other tracers, with consequences for the large-scale circulation. Although
Lagrangian power statistics do not directly quantify mixing or transport,
they provide a complementary means of examining the energy-exchange
dynamics experienced by moving fluid parcels. Examining how power
asymmetry varies with stratification and local flow structure may
therefore help clarify the Lagrangian dynamics of geophysical turbulence.

Previous studies have linked Lagrangian power statistics to energy
dissipation, local flow geometry, and pressure-mediated redistribution of
kinetic energy among fluid parcels
\cite{PumirredistributionPRX2014,PumirPRL2016vortexstretching,
picardo2020lagrangian}. Together, these connections indicate that power
asymmetry is associated not only with the overall state of turbulence but
also with the local dynamics encountered along particle trajectories.
Stratification adds another dimension to this problem by modifying the
balance of motions and structures within the flow. Irreversibility in
rotating-stratified turbulence has also been investigated through the
evolution of the relative kinetic energy of tracer pairs, offering a
complementary perspective based on particle-pair dynamics
\cite{Maity2019,Maity2020,Gallon2024}. However, how stratification affects power asymmetry
along individual particle trajectories, and whether the response differs
systematically among local flow structures, remain open questions.
Addressing these questions requires examining power statistics in relation
to local flow geometry rather than relying only on global statistics.

In this work, we investigate Lagrangian energy-exchange
statistics in direct numerical simulations of statistically stationary,
homogeneous Boussinesq turbulence across four dimensionless stratification
strengths. We first characterize the global distributions of kinetic,
potential, and total Lagrangian power, and examine how their fluctuation
amplitudes and standardized skewness vary with stratification. We then
investigate the directional dependence of kinetic-power asymmetry by
distinguishing horizontal and vertical component contributions. To relate these
statistics to the local flow structure, we classify sampled locations using the
invariants of the velocity-gradient tensor and examine power distributions
conditioned on four regions of the $Q$--$R$ plane. We further characterize the
persistence of tracer trajectories within these topological classes and examine
how their residence-time statistics vary with stratification. Finally, we
decompose kinetic power into pressure, viscous, buoyancy, and forcing
contributions, comparing their statistical properties and examining the
conditional pressure and buoyancy contributions to particle power. Together,
these analyses provide a broader characterization of how stratification
modifies Lagrangian power fluctuations, their dependence on local flow
structure, and the associated dynamical contributions. Throughout, we
distinguish statistical associations from causal mechanisms and emphasize that
Lagrangian power statistics provide a diagnostic of energy exchange rather than
a direct measurement of ocean mixing or transport.

\section{Methods and Governing Equations}

\subsection{Governing equations and numerical setup}

We investigate Lagrangian energy-exchange statistics in homogeneous turbulence governed by the incompressible Boussinesq equations. This framework describes a fluid in which density variations enter the dynamics through buoyancy, allowing us to examine how stable stratification modifies the velocity field and the motion of fluid parcels. The velocity field $\mathbf{u}$ and buoyancy field $b$ evolve according to
\begin{align}
\partial_t \mathbf{u}+\mathbf{u}\cdot\nabla\mathbf{u}
&=-\nabla P+\nu\nabla^2\mathbf{u}
+Nb\,\hat{\mathbf{z}}
+\mathbf{f},
\\
\partial_t b+\mathbf{u}\cdot\nabla b
&=\kappa\nabla^2 b
-Nu_z,
\end{align}
subject to the incompressibility condition
\begin{equation}
\nabla\cdot\mathbf{u}=0.
\end{equation}
Here, $P$ denotes the pressure variable in the momentum equation, $\nu$ is the kinematic viscosity, and $\kappa$ is the buoyancy diffusivity. The buoyancy field is denoted by $b$, $u_z$ is the vertical velocity component, and $\hat{\mathbf{z}}$ is the unit vector in the vertical direction. The Brunt--V\"ais\"al\"a frequency $N$ characterizes the strength of the stable background stratification. The forcing $\mathbf{f}$ sustains the turbulent flow against viscous dissipation.

The simulations are conducted in a triply periodic domain using a pseudospectral solver, dealiased with the 2/3 rule, with second-order integrating-factor Runge–Kutta (IFRK2) time integration and unit Prandtl number. The flow is maintained in a statistically stationary state by large-scale vortical forcing applied only to the horizontal velocity components, following the forcing protocol used in our previous work \cite{powdeluncertainty2026} and other studies of homogeneous stratified turbulence \cite{Varanasi2025, WAITE_BARTELLO_2004, Maffioli2016, Brethouwer2007}. This directional forcing is important when interpreting the results: the case $\widetilde N=0$ is referred to as the \emph{unstratified, but anisotropic, reference case}, and is therefore distinct from earlier studies in which the flow was isotropic.

The Taylor-scale Reynolds number is kept approximately constant across the
simulations, while the stratification strength is varied through the
dimensionless parameter $\widetilde N=N\tau_\eta$, where
$\tau_\eta=\left(\frac{\nu}{\varepsilon}\right)^{1/2}$ is the Kolmogorov time
scale and $\varepsilon$ is the mean kinetic-energy dissipation rate. This
parameterization allows us to compare the Lagrangian statistics across the
simulated stratification strengths while keeping the Reynolds number
approximately fixed.

The parameter $\widetilde N=N\tau_\eta$ compares the buoyancy time scale, $N^{-1}$, with the Kolmogorov time scale, $\tau_\eta$. Equivalently, for $N>0$, its inverse is the Kolmogorov-scale Froude number,
$Fr_\eta=(N\tau_\eta)^{-1}=\widetilde N^{-1}$,
which characterizes the relative importance of buoyancy and small-scale turbulent dynamics. We consider $\widetilde N=0$, $0.359$, $0.604$, and $0.999$, corresponding to $Fr_\eta$ values of approximately $2.79$, $1.66$, and $1.00$ for the three stratified cases, respectively. The unstratified case is treated separately because $Fr_\eta$ is undefined at $N=0$ and diverges in the limit $N\to0^+$. This parameter range allows us to examine changes in Lagrangian energy-exchange statistics from the unstratified reference case to a regime in which the buoyancy and Kolmogorov time scales become comparable \cite{BuariaPRF2020}. The condition $\widetilde N\sim1$ should be interpreted as a time-scale crossover rather than a sharp or universal boundary between distinct dynamical regimes.

To assess the geophysical relevance of the parameter range, we compare our simulations with representative oceanic conditions, following the approach used in our previous work \cite{powdeluncertainty2026}. Observations indicate Brunt--Väisälä frequencies in the range
$ \mathcal{O}(10^{-3}\,\mathrm{s}^{-1}) \lesssim N \lesssim \mathcal{O}(10^{-1}\,\mathrm{s}^{-1}) $
\cite{Emery1984,Millard,techreport2021}. Using representative oceanic values of the kinematic viscosity, $\nu=10^{-6}\,\mathrm{m}^2/\mathrm{s}$ \cite{Stanley1969}, and the mean kinetic-energy dissipation rate, $\varepsilon=10^{-6}\,\mathrm{m}^2/\mathrm{s}^3$ \cite{Lien2021}, gives a Kolmogorov time scale $\tau_\eta=(\nu/\varepsilon)^{1/2}\simeq1\,\mathrm{s}$. The corresponding dimensionless Brunt--Väisälä frequency, $\widetilde N=N\tau_\eta$, is therefore of order $10^{-3}$--$10^{-1}$. This overlaps with the lower part of the stratification range investigated here, which extends from $\widetilde N=0.359$ to $\widetilde N=0.999$, in addition to the unstratified reference case.

A complementary comparison can be made using the
integral-scale Froude number, $Fr=u_{\mathrm{rms}}/(\ell N)$, where
$u_{\mathrm{rms}}$ is the root-mean-square velocity and $\ell$ is the integral
length scale obtained from the kinetic-energy spectrum. Representative oceanic
values, $0.1\,\mathrm{m/s}\lesssim u_{\mathrm{rms}}\lesssim2\,\mathrm{m/s}$ and
$\mathcal{O}(10^4\,\mathrm{m})\lesssim\ell\lesssim\mathcal{O}(10^5\,\mathrm{m})$,
imply Froude numbers of order $10^{-4}$--$10^{-2}$ \cite{Jimenez}. This
provides a useful geophysical reference, although the integral-scale Froude
number and the Kolmogorov-scale Froude number characterize different scales and
should not be directly identified. Moreover, the Reynolds numbers accessible in
direct numerical simulations remain lower than those typical of the ocean.

The parameter range must also be interpreted in light of the
anisotropic forcing and the limitations of the simulations. In particular,
$\widetilde N$ alone does not determine whether the flow is
quasi-two-dimensional, since the influence of stratification can vary across
scales. For example, at $\widetilde N=0.359$, the Kolmogorov-scale Froude
number is approximately $2.79$, indicating that buoyancy and the smallest-scale
turbulent dynamics are not necessarily in the strongly stratified limit.
Moreover, the vortical forcing, which acts only on the horizontal velocity
components, can itself promote anisotropy at the forcing scales. The
simulations therefore provide a controlled comparison of Lagrangian statistics
across stratification strengths, but should not be interpreted as reproducing
all the features of oceanic turbulence.

\subsection{Lagrangian tracers and power statistics}

\begin{table*}[t]
	\caption{Mean $\mu$, standard deviation $\sigma$, and skewness $S$ of the Lagrangian power statistics. The columns report these three statistical moments for the kinetic power $p$, potential power $p_\mathrm{P}$, and total power $\pT=p+p_\mathrm{P}$, respectively.}
\label{tab:power_statistics}
\centering
\renewcommand{\arraystretch}{1.3}
\setlength{\tabcolsep}{9pt}
\begin{tabular}{c|ccc|ccc|ccc}
\hline\hline
& \multicolumn{3}{c}{$p$}
& \multicolumn{3}{c}{$\pP$}
& \multicolumn{3}{c}{$\pT$} \\
\cline{2-4} \cline{5-7} \cline{8-10}
$\widetilde{N}$
& $\mu$ & $\sigma$ & $S$
& $\mu$ & $\sigma$ & $S$
& $\mu$ & $\sigma$ & $S$ \\
\hline
$0$
& $0.00774$ & $14.2229$ & $-0.78652$
& -- & -- & --
& $0.00774$ & $14.2229$ & $-0.78652$ \\
$0.359$
& $0.00099$ & $8.23639$ & $-1.02199$
& $-0.00072$ & $2.55134$ & $-0.11791$
& $0.00026$ & $8.45504$ & $-0.93945$ \\
$0.604$
& $0.00606$ & $7.80227$ & $-1.74023$
& $0.00586$ & $2.75068$ & $-0.08728$
& $0.01192$ & $8.16457$ & $-1.56225$ \\
$0.999$
& $0.01295$ & $7.59499$ & $-2.32202$
& $-0.00071$ & $2.56884$ & $-0.44465$
& $0.01224$ & $8.05831$ & $-2.22922$ \\
\hline \hline
\end{tabular}
\end{table*}

To characterize energy changes along fluid trajectories, we seed tracer particles in the statistically stationary velocity field, with their positions $\mathbf{x}(t)$ evolving according to
\begin{equation}\label{eq:tracer_evolution_eq}
\frac{d\mathbf{x}}{dt}=\mathbf{u}(\mathbf{x}(t),t).
\end{equation}
The particles are passive: they follow the local fluid velocity without influencing the flow. Along each trajectory, we record, among other quantities, the velocity sampled by the tracer and define its instantaneous kinetic energy as
\begin{equation}
E(t)=\frac{1}{2}\left|\mathbf{u}(\mathbf{x}(t),t)\right|^2.
\end{equation}
The rate of change of this energy defines the instantaneous Lagrangian kinetic power,
\begin{equation}
p(t)=\frac{dE}{dt}.
\end{equation}
Positive power corresponds to an instantaneous gain of kinetic energy, whereas negative power corresponds to an energy loss. The distribution of $p$ therefore provides a statistical description of the kinetic-energy changes experienced by fluid parcels as they traverse the turbulent flow.

In stably stratified turbulence, buoyancy fluctuations also contribute to energy exchange. Represented by the buoyancy field $b$, these fluctuations allow us to define an instantaneous potential energy and its associated power along each trajectory:
\begin{equation}
E_{\mathrm{P}}(t)=\frac{1}{2} b(\mathbf{x}(t),t) ^2,
\qquad
p_{\mathrm{P}}(t)=\frac{dE_{\mathrm{P}}}{dt}.
\end{equation}
Combining the kinetic and potential contributions, we define the instantaneous total energy $E_{\mathrm{T}}=E+E_{\mathrm{P}}$ and total power $p_{\mathrm{T}}=p+p_{\mathrm{P}}$. The total power thus measures the instantaneous rate of change of the combined kinetic and potential energy along a fluid trajectory, whereas the individual powers characterize the changes in the two contributions separately. In a statistically stationary state, the mean total energy remains constant, implying $\langle p_{\mathrm{T}}\rangle=0$, as well as $\langle \pP \rangle =0 = \langle p \rangle$. This constraint does not imply that the instantaneous total power vanishes: kinetic and potential energy can fluctuate along trajectories, with nonlinear dynamics mediating their exchange. Moreover, the anisotropic forcing and stable stratification can break the equivalence of the Cartesian components of the velocity field, further influencing the statistics of Lagrangian energy changes.

We quantify the asymmetry of the power distributions using skewness. For any power observable $q\in\{p,p_{\mathrm{P}},p_{\mathrm{T}}\}$, its skewness is defined as
\begin{equation}
\label{eq:def_standardized_skewness}
S_q=
\frac{\left\langle(q-\langle q\rangle)^3\right\rangle}
{\left\langle(q-\langle q\rangle)^2\right\rangle^{3/2}}.
\end{equation}
For any observable $q$, one can define its standard deviation as $\sigma_q \equiv \langle(q -\langle q \rangle)^2\rangle^{1/2}$, so that Eq.\eqref{eq:def_standardized_skewness} reads $S_q = \langle(q - \langle q \rangle)^3\rangle/\sigma_q^3$. Here, angular brackets $\langle\cdots\rangle$ denote averages over time and over the ensemble of $N_p=2^{10}\times10^3$ Lagrangian trajectories. Negative values of $S_q$ indicate a distribution with a more pronounced negative tail, consistent with the flight--crash asymmetry commonly observed in turbulent energy fluctuations. Since skewness is a normalized third central moment, it characterizes the asymmetry of a distribution rather than its mean or overall amplitude. 

For comparison with finite-time measurements obtained from direct numerical simulations, we also consider energy increments over a time lag $\tau$, defined as $\delta_\tau \mathcal{E}=\mathcal{E}(t+\tau)-\mathcal{E}(t)$, where $\mathcal{E} \in \{ E, \EP, \ET \}$. 
In a statistically stationary flow, the mean energy increment vanishes, $\langle\delta_\tau \mathcal{E}\rangle=0$, because the mean energy does not change with time. For sufficiently small $\tau$, the increment satisfies $\delta_\tau \mathcal{E} \simeq p_\mathcal{E}(t)\tau$, where $p_\mathcal{E}(t)$ is the corresponding instantaneous power. Consequently, its standardized skewness approaches that of the instantaneous power as the time lag tends to zero, provided the relevant moments are well defined and the sampling resolves the associated time scales. In practice, we use energy increments with $\tau\ll\tau_\eta$, where $\tau_\eta$ is the Kolmogorov time scale, to estimate the small-time behavior of the power statistics. This approximation requires the time lag to be sufficiently small that the energy increment is dominated by the instantaneous rate of change; the resulting skewness can otherwise depend on the finite time lag. We therefore use power skewness as the primary diagnostic of Lagrangian energy-exchange asymmetry.

\section{Results}

\subsection{Global Statistics of Lagrangian Power}

\begin{figure}[t]
\centering
\includegraphics[width=\columnwidth]{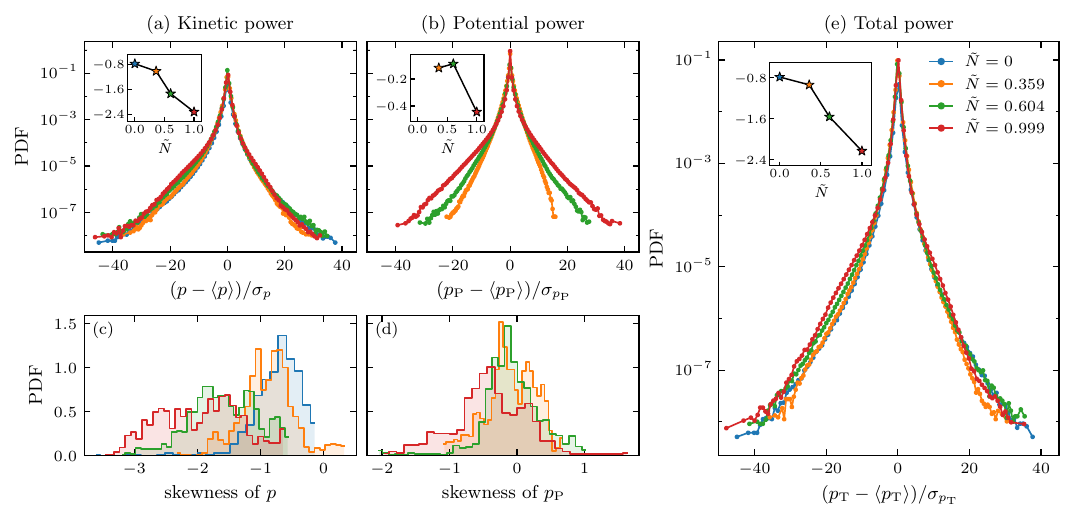}
\caption{Global statistics of Lagrangian power as a function of the stratification parameter $\widetilde{N}$. Panels (a), (b), and (e) show the probability distributions of the kinetic power $p$, potential power $\pP$, and total power $\pT$, respectively. The horizontal axes are shifted by their means and normalised by their standard deviations, $(q-\langle q \rangle)/\sigma_q$ with $q \in \{ p, \pP, \pT \}$ and $\sigma_q$ as defined below Eq.~\eqref{eq:def_standardized_skewness} to facilitate comparison of the changes in the distributions with increasing stratification. The insets show the corresponding skewness as a function of $\widetilde{N}$. Panels (c) and (d) show the distributions of snapshot-wise skewness values for the kinetic and potential power, respectively, in the statistically stationary state.}
\label{fig:power_pdfs}
\end{figure}

We begin by examining the global statistics of Lagrangian power, obtained by pooling samples over all tracer particles and sampling times in the statistically stationary state. The power at any instant is determined by the instantaneous state of the flow--particle system, although its statistical properties can reflect the system's prior evolution. Statistical stationarity implies that the probability distribution of power is invariant under shifts in the sampling time. Pooling samples across particles and times therefore provides a global characterization of energy-exchange statistics and allows us to compare their dependence on stratification. Two quantities are particularly relevant: the standard deviation, which measures the magnitude of fluctuations about the mean, and the skewness, which characterizes the asymmetry of the distribution.

The global statistics in Table~\ref{tab:power_statistics} reveal that these two measures respond differently to increasing stratification. Between the unstratified reference case, $\widetilde{N}=0$, and the most strongly stratified case, $\widetilde{N}=0.999$, the standard deviation of the kinetic power, $p$, decreases from 14.223 to 7.595, corresponding to a reduction of approximately $47\%$. Over the same range, its skewness becomes more negative, changing from $-0.787$ to $-2.322$. A similar trend is observed for the total power, $\pT$: its standard deviation decreases by approximately $43\%$, while its skewness changes from $-0.787$ to $-2.229$.

These results indicate that finite stratification is associated with weaker overall fluctuations in kinetic and total power but a more pronounced negative asymmetry in their distributions. Figure~\ref{fig:power_pdfs}(a,e) illustrates this behavior through the probability distributions of kinetic and total power at different stratification strengths. The increasingly pronounced negative tails are consistent with the observed decrease in skewness and confirm that the magnitude of power fluctuations and the asymmetry of their distribution need not vary in the same way.

The potential power exhibits a different response. Although its skewness remains negative across the stratification strengths considered, it does not show the systematic shift towards more negative values observed for the kinetic and total power. Thus, the enhancement of negative asymmetry in the total-power distribution is accompanied by a similar enhancement in the kinetic-power distribution, whereas the potential-power statistics do not display the same trend. This comparison motivates a closer examination of kinetic power as a diagnostic of Lagrangian energy-exchange asymmetry.

The global statistics describe the distribution obtained by pooling samples over particles and time, but they do not show how the skewness varies between individual snapshots in the sample space. To characterize this snapshot-to-snapshot variability, Fig.~\ref{fig:power_pdfs}(c,d) shows the distributions of skewness values calculated separately for each snapshot in the statistically stationary state. For kinetic power, the distribution shifts towards more negative skewness as stratification increases. By contrast, the corresponding distribution for potential power exhibits no comparable systematic trend. The agreement between the global and snapshot-wise kinetic-power statistics indicates that the increase in negative asymmetry with stratification is not merely a feature of pooling the samples over the full dataset.

These observations suggest that statistics based solely on the buoyancy
field, such as potential power, may not provide the most robust diagnostic
of Lagrangian time irreversibility in stratified turbulence.

The potential-power statistics nevertheless remain of interest in their own
right and warrant further investigation. Their physical interpretation,
however, requires care. For example, under a Galilean boost with velocity
$\mathbf{V}$, assuming the corresponding invariance of the Boussinesq
equations, the buoyancy field transforms as $b \to b' = b + NV_z t$.
Consequently, like the kinetic energy, the quantity $b^2/2$ is
frame-dependent and is meaningful in this formulation only in the frame in
which the mean vertical velocity vanishes, a condition that is also satisfied in
our simulations. Its interpretation is, however, less direct than that of
the kinetic energy associated with the velocity fluctuations. Furthermore,
buoyancy does not enter the tracer-position equation directly; its influence
on tracer motion is mediated by the velocity field. Taken together, these
considerations and the distinct trends observed in
Fig.~\ref{fig:power_pdfs} lead us not to pursue the potential power further
in this work.

On the other hand, the velocity field $\mathbf{u}$ considered here
represents the velocity fluctuations, with $\langle u_i\rangle = 0$ for
$i = x, y, z$. The kinetic energy $|\mathbf{u}|^2/2$ can therefore be used
directly to characterize their energy fluctuations. If the mean velocity
were non-zero, the corresponding fluctuation kinetic energy would instead be
defined using $|\mathbf{u} - \langle\mathbf{u}\rangle|^2/2$. We therefore
focus on the kinetic power in the remainder of this work.



\subsection{Directional Dependence of Power Asymmetry}

\begin{figure}[t]
\centering
\includegraphics[width=\columnwidth]{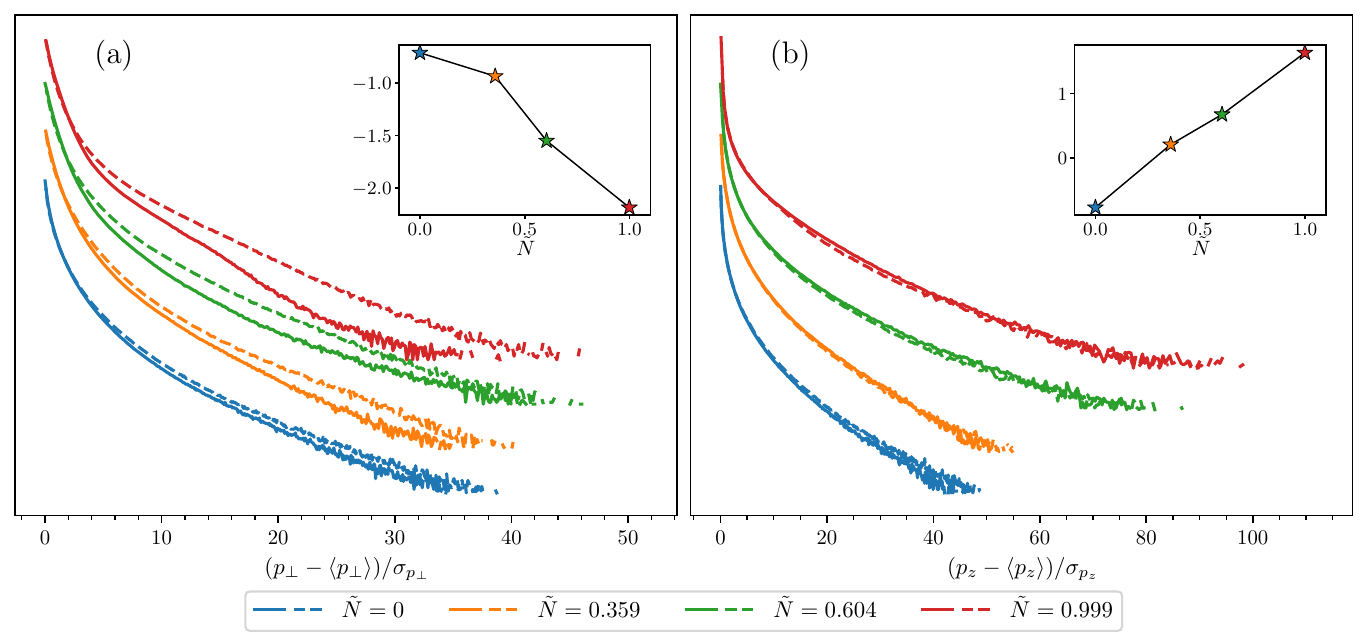}
\caption{Probability density functions (PDFs) of the (a) horizontal kinetic power $p_\perp$ and (b) vertical kinetic power $p_z$ for $\widetilde{N}=0$, $0.359$, $0.604$, and $0.999$ under anisotropic forcing. The abscissa shows $(q_i - \langle q_i \rangle)/\sigma_{q_i}$ where $q_i \in \{ p_\perp, p_z \}$ and $\sigma_{q_i}$ is the standard deviation as defined below Eq.~\eqref{eq:def_standardized_skewness}. Each distribution is folded about zero: solid lines represent the positive tail, while dashed lines represent the negative tail plotted against the absolute value of the power. Differences between the two tails therefore reveal the asymmetry of each distribution. For clarity, the curves are vertically offset by successive factors of $10$ as $\widetilde{N}$ increases; the ordinate is consequently shown in arbitrary units. The insets show the skewness of the corresponding power component as a function of $\widetilde{N}$. Both components are negatively skewed in the unstratified case. With increasing stratification, the skewness of $p_\perp$ becomes more negative, whereas that of $p_z$ increases and becomes positive.}
\label{fig:directional}
\end{figure}

We next examine the directional dependence of Lagrangian power asymmetry. In
homogeneous and isotropic turbulence, all spatial directions are statistically
equivalent, so the power associated with each velocity component, $u_i^2/2$ for
$i=x,y,z$, is expected to exhibit the same statistical properties, including
the negative skewness commonly associated with Lagrangian energy fluctuations.
In the present simulations, however, isotropy is broken even in the absence of
stratification because of the large-scale vortical forcing. To investigate how
this anisotropy and stable stratification influence Lagrangian time
irreversibility, we decompose the kinetic power into horizontal and vertical
contributions, $p=p_\perp+p_z$ where $p_\perp=p_x+p_y$, $p_x=u_x\frac{du_x}{dt}$,
$p_y=u_y\frac{du_y}{dt}$, and $p_z=u_z\frac{du_z}{dt}$.

Both $p_\perp$ and $p_z$ are Galilean-invariant observables in the present formulation, since $\mathbf{u}$ denotes the velocity fluctuations. Nevertheless, kinetic energy is a scalar, whereas the individual quantities $u_i^2/2$ depend on the choice of coordinate axes. We therefore use the componentwise powers as operational measures of directional energy changes rather than interpreting them as independent scalar measures of time irreversibility. Specifically, each componentwise power is defined as the product of the corresponding velocity and acceleration components, $p_i=u_i\,du_i/dt$.

Figure~\ref{fig:directional} shows that both horizontal and vertical power are negatively skewed in the unstratified case. Thus, despite the anisotropic forcing, the two components exhibit the same sign of skewness when stratification is absent. This observation indicates that the negative skewness of the componentwise power persists in the unstratified flow considered here, although it does not establish that the forcing scheme has no effect on its magnitude.

As stratification increases, the two components exhibit markedly different trends. The skewness of the horizontal power, $p_\perp$, becomes progressively more negative, indicating an increasingly pronounced negative asymmetry. In contrast, the skewness of the vertical power, $p_z$, increases with $\widetilde{N}$, evolving from negative values in the unstratified case to positive values at stronger stratification. The folded PDFs in Fig.~\ref{fig:directional} make this directional contrast visible: the relative prominence of the positive and negative tails changes differently for the horizontal and vertical components as stratification strengthens.

These results reveal that the increasingly negative skewness of the total kinetic power with increasing stratification masks distinct directional behaviors. The horizontal contribution develops stronger negative asymmetry, whereas the vertical contribution exhibits an increasingly positive asymmetry. Thus, the stratification-induced change in Lagrangian kinetic power statistics reflects distinct contributions of the different velocity components to the overall asymmetry, which the total kinetic power distribution alone does not reveal.

\subsection{How local flow topology relates to power asymmetry}

\begin{table*}[t]
\caption{Persistence time scale, $T_\mathrm{p}/\tau_\eta$, obtained from exponential fits to the tails of the persistence-time distributions conditioned on the four local velocity-gradient topology classes. The region definitions are provided in the discussion around Eqs.~\eqref{eq:def_Q_and_R}-\eqref{eq:persistence_fit}}.
\label{tab:persistence_decay}
\centering
\renewcommand{\arraystretch}{1.3}
\setlength{\tabcolsep}{14pt}
\begin{tabular}{l|l|c|c|c|c}
\hline
 & & \multicolumn{4}{c}{$\widetilde{N}$} \\
\cline{3-6}
Region & Condition & $0$ & $0.359$ & $0.604$ & $0.999$ \\
\hline
A: vortex compression  & $\Delta > 0,\ R > 0$ & $2.55$ & $2.06$ & $2.13$ & $2.12$ \\
B: vortex stretching & $\Delta > 0,\ R < 0$ & $4.54$ & $3.56$ & $3.58$ & $4.75$ \\
C: axial strain       & $\Delta < 0,\ R < 0$ & $1.37$ & $0.97$ & $0.85$ & $0.87$ \\
D: biaxial strain     & $\Delta < 0,\ R > 0$ & $2.90$ & $1.89$ & $1.66$ & $1.52$ \\
\hline
\end{tabular}
\end{table*}

\noindent
The local topology of the velocity-gradient tensor provides a framework for distinguishing regions with different instantaneous configurations of the flow \cite{chong1990, cantwell1993velgrad}. In particular, it separates regions in which the tensor has a complex-conjugate pair of eigenvalues, commonly associated with local rotation, from regions in which all three eigenvalues are real, associated with strain-dominated topology. This classification allows us to investigate how the asymmetry of Lagrangian power varies across local flow configurations and how these conditional statistics change with stratification.

Let $\mathcal{A} = \nabla\mathbf{u}$ denote the velocity-gradient tensor, with $\mathrm{Tr}(\mathcal{A}) = 0$ for an incompressible flow. We characterise its local topology using the second and third invariants,
\begin{equation}\label{eq:def_Q_and_R}
    Q = -\frac{1}{2}\mathrm{Tr}(\mathcal{A}^2), \qquad
    R = -\frac{1}{3}\mathrm{Tr}(\mathcal{A}^3),
\end{equation}
which map each sampled tracer position to the $Q$--$R$ plane. The discriminant
\begin{equation}\label{eq:def_discriminant}
    \Delta = \frac{27}{4}R^2 + Q^3
\end{equation}
distinguishes two broad eigenvalue configurations. For $\Delta > 0$, $\mathcal{A}$ has one real eigenvalue and a complex-conjugate pair, a configuration commonly associated with vorticity-dominated regions. For $\Delta < 0$, all three eigenvalues are real, corresponding to strain-dominated topology. This classification characterizes the instantaneous local velocity-gradient field; the energy transfer experienced by a tracer, including the sign of its instantaneous power, is therefore examined separately, conditioned on these classes.

We divide the $Q$--$R$ plane into four regions. \textbf{Region A}, defined by $\Delta > 0$ and $R > 0$, is conventionally termed the vortex-compression region, whereas \textbf{Region B}, with $\Delta > 0$ and $R < 0$, is termed the vortex-stretching region. The two remaining regions have $\Delta < 0$ and are associated with strain-dominated configurations. Since $R = -\det\mathcal{A}$, the sign of $R$ in these regions is determined by the signs of the three real eigenvalues. \textbf{Region C}, with $R < 0$, has one positive and two negative eigenvalues, corresponding to one stretching and two compressing directions, and is associated with axial strain. \textbf{Region D}, with $R > 0$, has two positive and one negative eigenvalue, corresponding to two stretching and one compressing direction, and is associated with biaxial strain. These labels summarize the geometry of the local velocity-gradient topology, capturing its dominant character rather than implying that the full local flow is pure strain, or by themselves determining the sign of particle power.

Before examining power statistics conditioned on these regions, we first consider how long tracers remain within each topology class. The persistence time, defined as the duration for which a tracer remains continuously within a given class, characterises the temporal sampling of local flow structures \cite{bhatnagar2016PRE,perlekar_persistence_2011}. It therefore provides complementary information to single-time statistics and helps establish whether the residence intervals associated with different topologies vary systematically with stratification.

\begin{figure}[t]
    \centering
    \includegraphics[width=\linewidth]{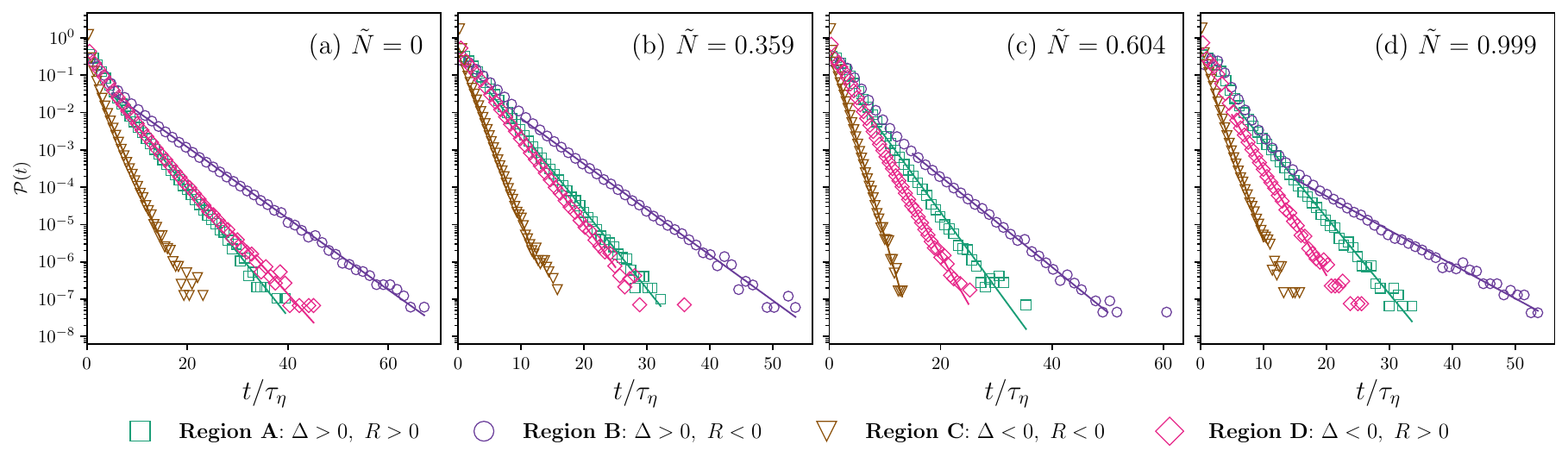}
    \caption{Probability distributions of Lagrangian persistence times in the four velocity-gradient topology classes at (a) $\widetilde N = 0$, (b) $\widetilde N=0.359$, (c) $\widetilde N=0.604$, and (d) $\widetilde N=0.999$. The persistence time is the duration for which a tracer remains continuously within a given class. Solid curves denote exponential fits of the form $\mathcal{P}(t)=\mathcal{C} \exp[-t/T_p]$. The fitted persistence time scales are reported in Table~\ref{tab:persistence_decay}. Differences in the fitted decay scales indicate how the distribution of residence intervals varies across topology classes and stratification strengths.}
    \label{fig:persistence}
\end{figure}

Figure~\ref{fig:persistence} shows the probability distributions of persistence times for the four topology classes at $\widetilde{N}=0$, $0.359$, $0.604$, and $0.999$, together with exponential fits of the form
\begin{equation}
    \mathcal{P}(t) = \mathcal{C} \exp\left(-\frac{t}{T_p}\right),
    \label{eq:persistence_fit}
\end{equation}
where $\mathcal{C}$ is the fitted prefactor and $T_p$ is the persistence time scale. The fitted values of $T_p$ are reported in Table~\ref{tab:persistence_decay}.

A notable feature is the crossover between the persistence-time scales of Regions A and D as stratification increases. In the unstratified case, $T_p/\tau_\eta$ is larger for Region D than for Region A. At nonzero stratification, this ordering reverses, and the difference between the two scales becomes larger. Thus, the relative persistence of these two topological classes changes with stratification.

The fitted distributions indicate that long residence intervals become less prevalent with increasing stratification in Regions A, C, and D, whereas Region B exhibits the opposite trend over the sampled range.

These differences show that stratification alters the temporal sampling of local flow topologies, although the changes are not uniform across the four classes. However, persistence statistics alone do not establish whether the observed changes in global Lagrangian power asymmetry arise from changes within individual classes or from differences in how tracers sample those classes. To address this distinction, we next condition the power statistics on the four regions of the velocity-gradient topology defined above.

\begin{figure}[t]
\centering
    \includegraphics[width=\columnwidth]{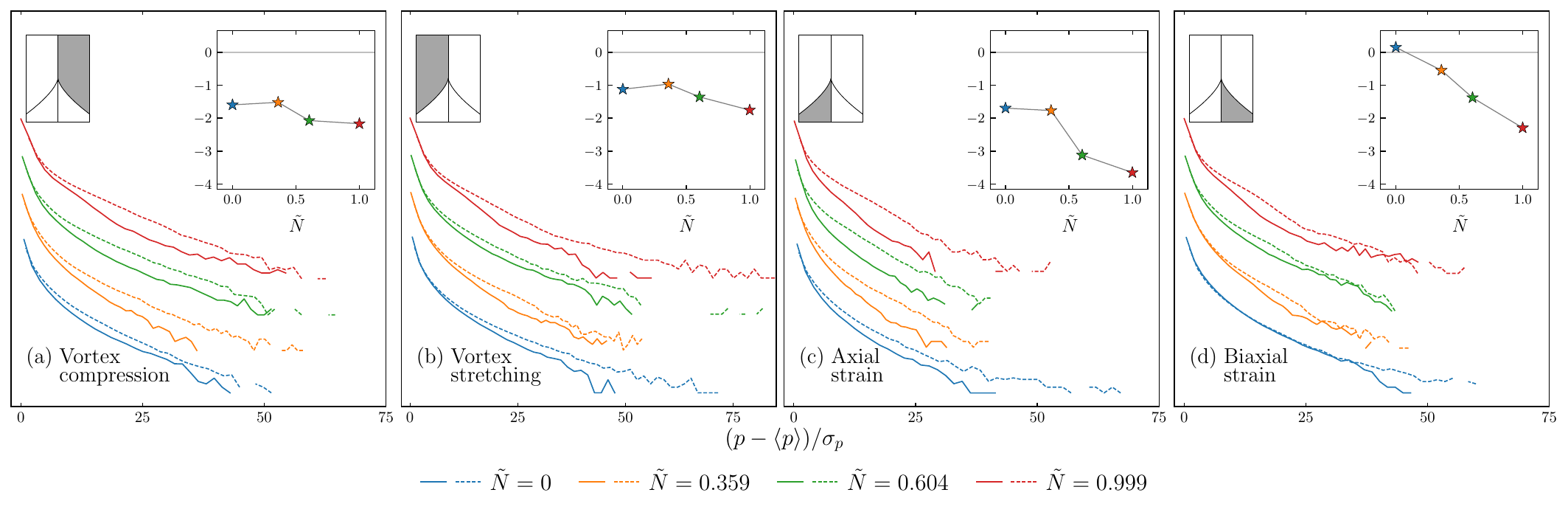}
    \caption{Probability density functions (PDFs) of shifted and normalized Lagrangian power $p$, conditioned on the local flow topology in each of the four regions of the $Q$--$R$ plane, for $\widetilde{N}=0$, $0.359$, $0.604$, and $0.999$. The averages in the mean, $\langle p \rangle$ and standard deviation, $\sigma_p$ are also conditioned on local flow topology. The shaded area in the small diagram in each panel marks the region concerned. Regions A and B lie above the zero-discriminant curve ($\Delta>0$) and are vorticity dominated: A ($R>0$) corresponds to vortex compression, and B ($R<0$) to vortex stretching. Regions C and D ($\Delta<0$) are strain dominated: C ($R<0$) corresponds to axial strain, with one stretching and two compressing directions, while D ($R>0$) corresponds to biaxial strain, with two stretching directions and one compressing direction. Solid lines show the positive tail of each PDF, and dashed lines show the negative tail reflected about zero; for clarity, successive curves are offset vertically by factors of $10^2$. Insets show the skewness of $p$ in each region as a function of $\widetilde{N}$. These topological classes describe the local velocity-gradient field and do not, by themselves, determine the sign of a particle's instantaneous power.}
    \label{fig:topology}
\end{figure}

Table~\ref{tab:topology_skewness} reports the skewness of the total, perpendicular, and vertical kinetic power conditioned on each of these regions.

Figure~\ref{fig:topology} shows the PDFs of Lagrangian power conditioned on each topological class, together with the corresponding skewness as a function of stratification. The conditional distributions reveal substantial differences in the asymmetry of power fluctuations across the four regions.

Region C (axial strain) is the most negatively skewed of the four classes at every sampled value of $\widetilde{N}$. Its power distribution therefore exhibits the strongest negative skewness throughout the parameter range examined. Region D (biaxial strain) behaves differently: its skewness is weakly positive in the unstratified case ($0.153$) but becomes negative at all three stratified values, reaching $-2.290$ at $\widetilde{N}=0.999$. Its skewness therefore changes sign between $\widetilde{N}=0$ and $\widetilde{N}=0.359$; the present samples do not locate the crossing more precisely.

Regions A (vortex compression) and B (vortex stretching) are negatively skewed throughout. In both regions, the skewness first becomes slightly less negative at the lowest nonzero stratification and then becomes more negative as $\widetilde{N}$ increases further. Their response is therefore non-monotonic rather than a steady strengthening of negative asymmetry.

Taken together, these conditional statistics demonstrate that stratification does not modify the asymmetry of Lagrangian power uniformly across local velocity-gradient structures. Both the magnitude of the skewness and its evolution with $\widetilde{N}$ depend on the topology class. The contrast between Regions C and D is particularly pronounced: Region C retains strong negative skewness throughout the sampled range, whereas Region D changes from weakly positive skewness in the unstratified case to pronounced negative skewness under stratification. Consequently, global power statistics conceal substantial heterogeneity in the conditional response of different local flow structures.

The skewness of a conditional power distribution quantifies its asymmetry, but does not identify the dynamical contributions responsible for that asymmetry. Moreover, the persistence statistics show that tracers do not spend the same characteristic time in every topology class, and these residence times respond differently to stratification. Interpreting the global statistics therefore requires distinguishing changes in power fluctuations within each class from changes in the temporal sampling of the classes. The dynamical decomposition of power provides an additional means of examining these conditional differences.

\begin{table*}[t]
\caption{Standardized skewness of kinetic Lagrangian power $p$, horizontal (perpendicular) component of kinetic power $p_\perp$, and vertical kinetic power $p_z$, conditioned on the four local velocity-gradient topology classes. The definitions of the regions are provided in the discussion around Eqs.~\eqref{eq:def_Q_and_R}-\eqref{eq:persistence_fit}.}
\label{tab:topology_skewness}
\centering
\renewcommand{\arraystretch}{1.3}
\setlength{\tabcolsep}{5pt}
\begin{tabular}{c|ccc|ccc|ccc|ccc}
\hline
& \multicolumn{3}{c}{Region A}
& \multicolumn{3}{c}{Region B}
& \multicolumn{3}{c}{Region C}
& \multicolumn{3}{c}{Region D} \\
& \multicolumn{3}{c}{vortex compression}
& \multicolumn{3}{c}{vortex stretching}
& \multicolumn{3}{c}{axial strain}
& \multicolumn{3}{c}{biaxial strain} \\
\cline{2-4} \cline{5-7} \cline{8-10} \cline{11-13}
$\widetilde{N}$
& $p$ & $p_\perp$ & $p_z$
& $p$ & $p_\perp$ & $p_z$
& $p$ & $p_\perp$ & $p_z$
& $p$ & $p_\perp$ & $p_z$ \\
\hline
$0$
& $-1.593$ & $-1.402$ & $-1.441$
& $-1.118$ & $-0.757$ & $-0.975$
& $-1.691$ & $-1.466$ & $-1.694$
& $0.153$ & $0.070$ & $0.440$ \\
$0.359$
& $-1.515$ & $-1.545$ & $-0.322$
& $-0.963$ & $-0.936$ & $0.620$
& $-1.765$ & $-1.540$ & $-0.836$
& $-0.543$ & $-0.286$ & $0.172$ \\
$0.604$
& $-2.068$ & $-2.105$ & $0.034$
& $-1.352$ & $-1.379$ & $1.827$
& $-3.117$ & $-2.698$ & $-2.623$
& $-1.375$ & $-0.912$ & $0.261$ \\
$0.999$
& $-2.165$ & $-2.273$ & $1.487$
& $-1.751$ & $-2.164$ & $4.271$
& $-3.646$ & $-3.104$ & $-2.969$
& $-2.290$ & $-1.755$ & $-0.524$ \\
\hline
\end{tabular}
\end{table*}

\subsection{Dynamical Contributions to Lagrangian Power}

The topology-conditioned statistics show that the asymmetry of Lagrangian power varies substantially among local velocity-gradient classes and changes with stratification. To investigate the dynamical origins of this asymmetry, we return to the Boussinesq momentum equation and decompose the instantaneous rate of change of kinetic energy along a tracer trajectory into pressure, viscous, buoyancy, and forcing contributions:
\begin{equation}
\label{eq:power_evolution}
p = \mathbf{u}\cdot\frac{D\mathbf{u}}{Dt}
= - \mathbf{u}\cdot\nabla P
+ \nu \mathbf{u}\cdot\nabla^2\mathbf{u}
+ N b u_z
+ \mathbf{u}\cdot\mathbf{f}.
\end{equation}
Here, $D/Dt$ denotes the material derivative, $P$ is the pressure variable appearing in the momentum equation, $\nu$ is the kinematic viscosity, $b$ is the buoyancy variable, $N$ is the buoyancy frequency, and $\mathbf{f}$ denotes the forcing. The decomposition identifies the instantaneous physical contributions to kinetic power along a tracer trajectory.

The third central moment of the total power, $\langle (p-\langle p\rangle)^3\rangle$, quantifies its asymmetry. However, because $p$ is a sum of several contributions, its third central moment depends not only on the individual third central moments but also on mixed moments involving different contributions. Consequently, the third central moment of an individual term cannot be interpreted as its additive contribution to the third central moment of the total power. We therefore compare the third central moments of the individual terms with that of the total power, using this comparison to characterize their respective statistical asymmetries rather than to construct an additive partition of the total asymmetry.

Among the terms in Eq.~\eqref{eq:power_evolution}, the pressure-gradient contribution is distinguished by its non-local character. The pressure field is constrained by a Poisson equation involving the velocity-gradient field and buoyancy, so its spatial variation reflects the surrounding flow rather than only the instantaneous local state. Previous studies of unstratified turbulence have shown that, despite having a vanishing mean under appropriate statistical conditions, pressure plays an important role in redistributing kinetic energy among fluid particles \cite{PumirredistributionPRX2014}. The statistical role of pressure has also been shown to differ markedly between two- and three-dimensional flows. Since stratification can promote two-dimensionalization, it is therefore interesting to examine whether the pressure contribution develops statistical features associated with both three-dimensional and two-dimensional turbulence as stratification increases.

Table~\ref{tab:dynamic_contributions} reports the mean, standard deviation, and standardized skewness of the total kinetic power and of the individual dynamical contributions in Eq.~\eqref{eq:power_evolution}, together with the third central moments of the individual contributions normalized by the third central moment of the total kinetic power (we hereby, refer to it the normalized third central moment). The standard deviation characterizes the amplitude of the fluctuations of each contribution, whereas its standardized skewness measures the asymmetry of its distribution independently of its overall amplitude. These quantities therefore provide complementary information: a contribution can exhibit substantial fluctuations without necessarily having a strongly asymmetric distribution, and a large magnitude of skewness does not by itself imply a large contribution to the overall power fluctuations. The normalized third central moments compare the signs and relative magnitudes of the individual third moments with that of the total power. However, they do not quantify the additive contributions of the terms to the total third moment because the mixed moments are not included.

The pressure contribution has a standard deviation comparable to that of the total kinetic power, decreasing from $14.1$ in the unstratified case to $7.57$ at $\widetilde{N}=0.999$. Its standardized skewness changes substantially with stratification: it is nearly zero in the unstratified case ($S_P=0.00524$) and becomes increasingly negative, reaching $-1.56$ at the largest stratification considered. Thus, the pressure contribution develops a pronounced negative asymmetry while its fluctuation amplitude decreases. Its normalized third central moment also changes from a small negative value ($-0.0065$) in the unstratified case to positive values in all three stratified cases, increasing from $0.287$ to $0.666$. This behavior is noteworthy in light of previous findings that pressure-related statistics differ between three-dimensional and two-dimensional turbulence \cite{PumirredistributionPRX2014}.

The viscous contribution has a substantially smaller standard deviation than the pressure and total-power contributions, with values between $1.70$ and $2.59$. Its standardized skewness is strongly negative throughout the parameter range, varying from $-5.30$ in the unstratified case to values between $-3.41$ and $-3.93$ in the stratified cases. Despite this pronounced standardized asymmetry, its normalized third central moment is comparatively small and decreases from $0.0408$ to $0.0191$. This illustrates why skewness alone should not be used to assess the relative magnitude of a contribution's third central moment: the latter also depends on the cube of the standard deviation.

The direct buoyancy contribution, which is absent in the unstratified case, has a standard deviation between $2.38$ and $2.64$ across the stratified cases. Its standardized skewness remains negative, ranging from $-0.888$ to $-1.18$, while its normalized third central moment remains comparatively small, between $0.0186$ and $0.0221$. Thus, although the buoyancy contribution exhibits a measurable asymmetry, its third central moment is small relative to that of the total kinetic power.

The change in the pressure contribution raises the question of whether it signals a transition towards quasi-two-dimensional dynamics. However, the case $\widetilde{N}=0.359$ is not necessarily quasi-two-dimensional: at this stratification strength, the Kolmogorov-scale Froude number is $\text{Fr}_\eta=2.79$, and stratification has not yet influenced all scales of the flow. The observed pressure statistics may therefore reflect an interplay between the three-dimensional dynamics and the tendency towards two-dimensionalization. The vortical nature of the forcing may also influence the dynamics at the forcing scales. Nevertheless, the present statistics do not establish a causal connection between either mechanism and the change in the pressure contribution. Determining how this behavior depends on the forcing would require further investigation, including simulations with isotropic forcing.

Taken together, these statistics distinguish the fluctuation amplitudes and standardized asymmetries of the individual dynamical contributions from their third central moments relative to that of the total power. The pressure contribution exhibits the most pronounced change with stratification in both its standardized skewness and normalized third central moment. However, neither the standardized skewness nor the normalized third-moment ratios provide an additive partition of the total-power asymmetry. A quantitative attribution would also require the mixed third-order moments between the contributions in Eq.~\eqref{eq:power_evolution}.

\begin{table*}[t]
\caption{Mean, standard deviation, and standardized skewness of the total power $p$ and of each term contributing to $p$ in Eq.~\eqref{eq:power_evolution}. To facilitate comparison with the asymmetry of $p$, the ratio of the third central moment of each term to that of $p$ is also shown. The buoyancy term is absent in the unstratified case. Note that $\sigma_p^{(b)} =  \langle [ (u_z b) - \langle u_z b \rangle ]^2 \rangle^{1/2}$. }
\label{tab:dynamic_contributions}
\centering
\renewcommand{\arraystretch}{1.3}
\setlength{\tabcolsep}{8pt}
\begin{tabular}{c|c|c|c|c}
\hline\hline
Quantity & $\widetilde{N} = 0$ & $\widetilde{N} = 0.359$ & $\widetilde{N} = 0.604$ & $\widetilde{N} = 0.999$ \\
\hline
\textit{Total kinetic power} & & & & \\
$\langle p \rangle$
  & $7.74 \times 10^{-3}$ & $9.9 \times 10^{-4}$ & $6.06 \times 10^{-3}$ & $1.30 \times 10^{-2}$ \\
$\sigma_p$
  & $14.2$ & $8.24$ & $7.80$ & $7.59$ \\
$S_p$
  & $-0.787$ & $-1.02$ & $-1.74$ & $-2.32$ \\
\hline
\textit{Pressure contribution to power} & & & & \\
$\langle -\mathbf{u} \cdot \nabla P \rangle$
  & $-2.74 \times 10^{-4}$ & $-3.02 \times 10^{-5}$ & $-6.88 \times 10^{-5}$ & $7.76 \times 10^{-4}$ \\
  $\sigma_p^{(P)}$ & 14.1 & 8.17 & 7.69 & 7.57 \\
$S_p^{(P)}$ & $0.00524$ & $-0.301$ & $-0.715$ & $-1.56$ \\
$\langle [ -\mathbf{u} \cdot \nabla P - \langle -\mathbf{u} \cdot \nabla P \rangle ]^3 \rangle / \langle (p - \langle p \rangle)^3 \rangle$
  & $-0.0065$ & $0.287$ & $0.394$ & $0.666$ \\
\hline
\textit{Viscous contribution to power} & & & & \\
$\langle \nu\, \mathbf{u} \cdot \nabla^2 \mathbf{u} \rangle$
  & $-0.898$ & $-0.630$ & $-0.676$ & $-0.713$ \\
$\sigma_p^{(\nu)}$   & $2.59$    & $1.71$   & $1.83$   & $1.70$ \\
$S_p^{(\nu)}$        & $-5.30$   & $-3.41$  & $-3.85$  & $-3.93$ \\
$\langle [ \nu\, \mathbf{u} \cdot \nabla^2 \mathbf{u} - \langle \nu\, \mathbf{u} \cdot \nabla^2 \mathbf{u} \rangle ]^3 \rangle / \langle (p - \langle p \rangle)^3 \rangle$
  & $0.0408$ & $0.0300$ & $0.0286$ & $0.0191$ \\
\hline
\textit{Buoyancy contribution to power} & & & & \\
$\langle N b u_z \rangle$
  & -- & $-0.269$ & $-0.218$ & $-0.176$ \\
$N \sigma_p^{(b)}$  & --        & $2.38$   & $2.64$   & $2.52$ \\
$S_p^{(b)}$       & --        & $-0.888$ & $-0.990$ & $-1.18$ \\
$\langle [ N b u_z - \langle N b u_z \rangle ]^3 \rangle / \langle (p - \langle p \rangle)^3 \rangle$
  & -- & $0.0208$ & $0.0221$ & $0.0186$ \\
\hline\hline
\end{tabular}
\end{table*}

Overall, this decomposition identifies the pressure contribution as the term whose normalized third central moment changes most substantially with stratification, while the viscous and direct buoyancy contributions remain comparatively small in magnitude. These results highlight a change in the statistical asymmetry of the individual dynamical contributions, but do not by themselves establish which terms dominate the asymmetry of total kinetic power. A quantitative attribution would also require the mixed third-order moments between the contributions in Eq.~\eqref{eq:power_evolution}.

\subsection{Pressure-mediated Energy Redistribution and Buoyancy Contributions}

\begin{figure}[t]
\centering
\includegraphics[width=\columnwidth]{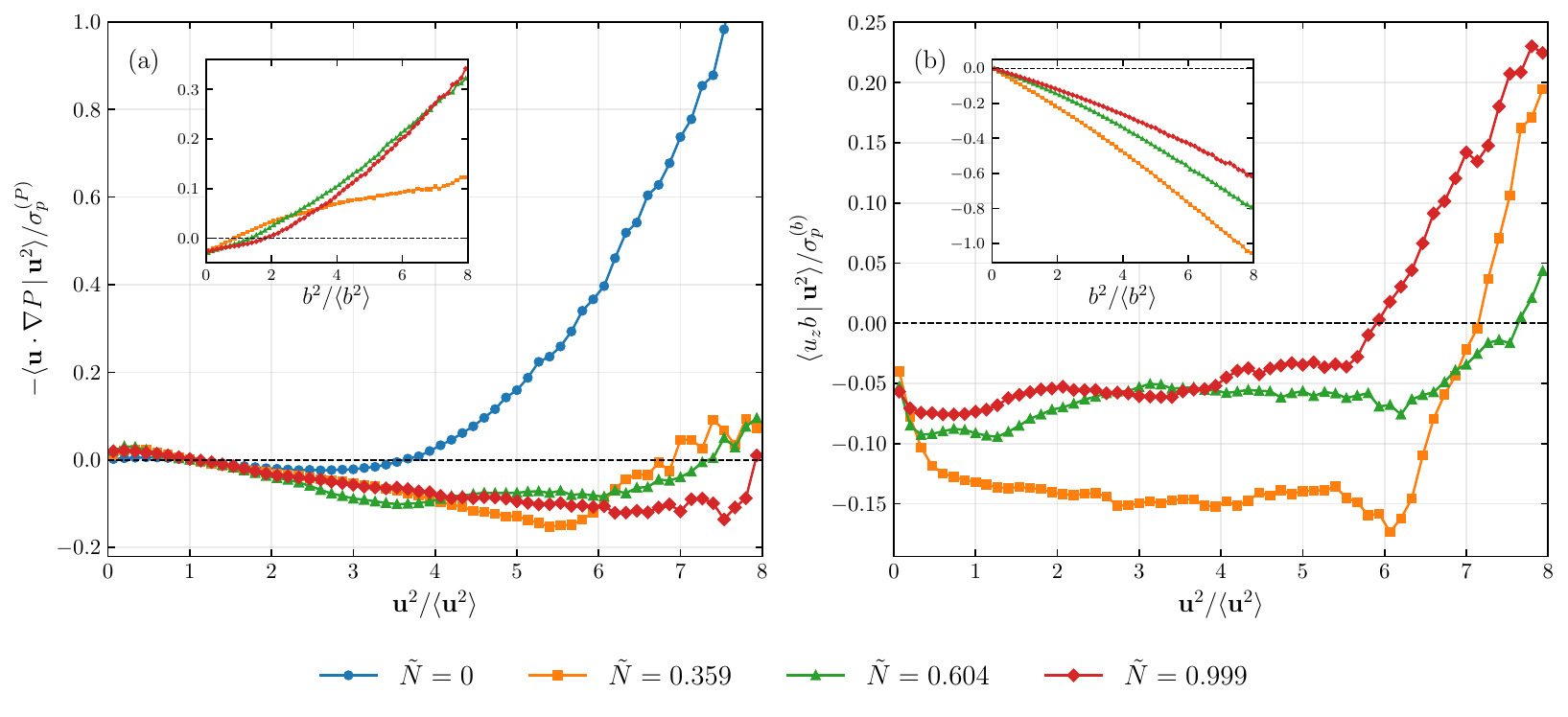}
\caption{Conditional averages of the pressure and buoyancy contributions to particle power. (a) Pressure contribution, $-\langle \mathbf{u}\cdot\nabla P \,|\, \mathbf{u}^2\rangle$, normalized by its standard deviation, $\sigma_p^{(P)} = \langle [ -\mathbf{u} \cdot \nabla P - \langle -\mathbf{u} \cdot \nabla P \rangle ]^2 \rangle^{1/2} $ and conditioned on normalized particle kinetic energy, $\mathbf{u}^2/\langle \mathbf{u}^2\rangle$. (b) Buoyancy contribution, $\langle u_z b \,|\, \mathbf{u}^2\rangle$, normalized by $\sigma_p^{(b)} = \langle [ (u_z b) - \langle u_z b \rangle ]^2 \rangle^{1/2}$ and conditioned on normalized particle kinetic energy. The buoyancy contribution vanishes in the absence of stratification, so the $\widetilde{N}=0$ case is not shown. Insets show the corresponding conditional averages in the stratified runs as functions of normalized squared buoyancy, $b^2/\langle b^2\rangle$, which characterizes the particle's potential-energy fluctuations. In the unstratified case, pressure contributes negatively to the power of particles with kinetic energies approximately between $\langle \mathbf{u}^2\rangle$ and $4\langle \mathbf{u}^2\rangle$, while its contribution becomes strongly positive for the most energetic particles. With stratification, pressure contributes positively for slower particles and negatively over a broad range of higher kinetic energies, consistent with a restoring tendency in the conditional mean over that range. The pressure contribution also becomes positive for particles with large $b^2$, as shown in the inset of (a). The buoyancy contribution is negative over most of the sampled kinetic-energy range, indicating a mean transfer out of kinetic energy over that range. Its negative contribution becomes more pronounced with increasing $b^2$ in the inset of (b), and is strongest in normalized terms for the weakest nonzero stratification.}
\label{fig:pressure_avg}
\end{figure}

\noindent
The preceding analysis identified the pressure contribution as particularly interesting because its third-moment statistics change substantially with stratification. We now examine this contribution more directly by conditioning the instantaneous pressure power on particle kinetic energy. This approach allows us to compare the mean pressure contribution across particles with different kinetic energies and thereby complements the third-moment analysis. We also examine the corresponding conditional statistics of the buoyancy contribution to determine how it varies with particle kinetic energy and buoyancy amplitude.

The pressure field is not an independent dynamical variable; its gradient is constrained by a Poisson equation,
\begin{equation}
\nabla^{2}P
=
-\partial_i u_j\,\partial_j u_i
+
N\,\partial_z b,
\label{eq:pressure}
\end{equation}
whose source depends on the velocity-gradient tensor and the buoyancy field. Pressure therefore reflects the surrounding flow as well as the local velocity-gradient configuration. Stable stratification reorganizes these fields and is accompanied by changes in the conditional pressure contribution to particle power. The statistics examined below provide evidence of stratification-dependent pressure-mediated energy redistribution, although they do not establish a unique causal pathway linking this redistribution to the observed asymmetry of Lagrangian power.

Figure~\ref{fig:pressure_avg}(a) shows the conditional mean pressure contribution to instantaneous power as a function of normalized particle kinetic energy, $\mathbf{u}^{2}/\langle\mathbf{u}^{2}\rangle$. The dependence changes markedly between the unstratified reference case and the stratified cases. With stratification, pressure contributes positively to the mean power of slower particles, for which $\mathbf{u}^{2}/\langle\mathbf{u}^{2}\rangle\leq 1$, and negatively over a broad range of higher particle energies. At the largest kinetic energies shown, the conditional trend moves back towards the behavior observed in the unstratified case. These results indicate that the mean pressure contribution depends on particle kinetic energy and that this dependence changes with stratification.

Figure~\ref{fig:pressure_avg}(b) shows the conditional mean buoyancy contribution to kinetic power. This contribution is negative over much of the sampled kinetic-energy range, indicating that buoyancy removes kinetic energy on average from particles in those bins. In a stably stratified flow, this negative contribution is consistent with kinetic energy being transferred into potential energy. The corresponding conditional averages in the insets of Fig.~\ref{fig:pressure_avg} show how the pressure and buoyancy contributions vary with normalized squared buoyancy, $b^2/\langle b^2\rangle$. The pressure contribution becomes positive at large values of this variable, whereas the buoyancy contribution becomes more negative. These trends characterize the conditional means and reveal a clear statistical association between a particle’s buoyancy amplitude and its subsequent energy exchange, whose causal nature remains to be established.

Over the kinetic-energy range containing most of the probability, the conditional pressure statistics are consistent with a restoring tendency: pressure tends to increase the power of slower particles and decrease it across a broad range of faster particles. This interpretation applies to the conditional mean, not to every individual trajectory, and does not establish that pressure acts as a restoring force in a dynamical or causal sense. Read together with the third-moment statistics discussed above, these results support an association between pressure-mediated energy redistribution and the stratification-dependent asymmetry of Lagrangian power. The conditional buoyancy statistics further show that the direct buoyancy contribution is predominantly negative over much of the sampled kinetic-energy range. Determining how these conditional trends combine to produce the global power asymmetry would require additional analysis of the joint statistics and cross-correlations among the dynamical contributions.

\section{Conclusion}

We investigated Lagrangian power asymmetry in homogeneous stratified turbulence, focusing on its dependence on local velocity-gradient topology, directional motion, and the temporal sampling of flow structures. Across the sampled stratification range, the variance of kinetic power decreases while its standardized skewness becomes more negative. Thus, the overall fluctuation level and the asymmetry of particle energy exchange exhibit distinct responses to stratification, underscoring the importance of considering both measures when characterizing Lagrangian power statistics.

Conditioning the power statistics on local velocity-gradient topology reveals substantial differences that are obscured by global averages. The clearest contrast occurs between the two strain-dominated classes. Region C, associated with axial strain, exhibits the most negative total-power skewness at every sampled stratification strength. Region D, associated with biaxial strain, changes from positive skewness in the unstratified reference case to negative skewness in all three stratified cases. Regions A and B, associated with vortex compression and vortex stretching, respectively, also exhibit negative skewness, with a non-monotonic response at the lowest nonzero stratification. These results demonstrate that power asymmetry varies substantially across local velocity-gradient configurations, although the conditional statistics alone cannot establish that topology causes the observed asymmetry.

The horizontal and vertical component-power statistics reveal further directional differences as stratification increases, consistent with the anisotropic response of the flow. The dynamical decomposition provides a complementary perspective on these changes. The normalized third central moment of the pressure contribution increases markedly with stratification, whereas that of the viscous contribution decreases; the direct buoyancy contribution remains comparatively small in magnitude. These individual moments are not additive components of the total third central moment, because mixed moments between the dynamical contributions also enter the latter. The energy-conditioned statistics additionally show that the mean pressure contribution tends to increase the power of slower particles and decrease it over a broad range of higher particle energies. Taken together, these observations associate the stratification-dependent power statistics with changes in pressure-mediated energy redistribution, without identifying a unique causal mechanism.

The persistence-time distributions provide a complementary temporal perspective. Their tails are described by exponential fits, with fitted persistence time scales that vary across topology classes and respond differently to stratification. The persistence time scales decrease with increasing stratification in Regions A, C, and D, whereas Region B exhibits an increase at the largest stratification considered. These differences indicate that the temporal sampling of local flow structures is not uniform across topology classes. Persistence statistics therefore complement instantaneous power statistics by characterizing how long tracers remain within each class. However, the present analysis does not establish a direct relationship between residence duration and power asymmetry, nor does it determine how residence times and class occupancy combine to shape the global statistics.

Overall, our results show that Lagrangian power asymmetry in stratified turbulence varies with local flow topology, differs between horizontal and vertical motion, and accompanies changes in pressure-mediated energy redistribution and the temporal sampling of flow structures. These conclusions are restricted to the simulated parameter range and forcing configuration; the diagnostics presented here do not directly quantify ocean mixing or transport. An important direction for future work is to combine topology-conditioned power statistics with tracer residence histories and subsequent energy increments. Such an analysis could test whether the duration and sequence of visits to different topological classes are statistically associated with subsequent energy exchange, helping clarify how temporal sampling relates to the power asymmetry observed along Lagrangian trajectories.

\begin{acknowledgments} 
     M.J.P. thanks Rama Govindarajan for useful comments. S.S.R. acknowledges the Indo-French Centre for the
Promotion of Advanced Scientific Research (IFCPAR/CEFIPRA, Project No. 6704-1) for support. This research was supported in part by the International Centre
for Theoretical Sciences (ICTS) for the program --- 11th Indian Statistical
Physics Community Meeting (code: ICTS/11thISPCM2026/04). 
The simulations were performed on the ICTS clusters Mario, Tetris, and Contra. M.J.P. and
S.S.R. acknowledge the support of the DAE, Government of India, under Projects No. RTI4019 and
No. RTI4013.
\end{acknowledgments}

\bibliography{references}
\end{document}